\documentclass[9pt,twocolumn]{revtex4}
\usepackage{graphicx}
\usepackage{amssymb}
\usepackage{amsmath}
\usepackage{bm}
\usepackage{color}

\begin{document}

\title{Ballistic and shift currents in the bulk photovoltaic effect theory}

\author{Boris Sturman}
\affiliation{Institute of Automation and Electrometry, Russian Academy of Sciences,
Koptyug ave., 630090 Novosibirsk, Russia
 }

\begin{abstract}
The bulk photovoltaic effect (BPVE) -- generation of electric currents by light in
noncentrosymmetric materials in the absence of electric fields and gradients  -- has
been intensively investigated in the end of the last century. The outcomes including all
main aspects of this phenomenon were summarized in review and books. A new upsurge of
interest to the BPVE occurred recently and resulted in a flux of misleading theoretical
and experimental publications centered around the so-called shift current. Numerous
top-rated recent publications ignore the basic principles of charge-transport phenomena
and the previous results of joined experimental-theoretical studies. Specifically,
dominating (or substantial) contributions to the currents caused by asymmetry of the
momentum distributions of electrons and holes are missed. The widespread starting
relation for the shift current, originating from the quadratic nonlinear response theory
and pretending to be omnipotent, is in fact incomplete. It ignores kinetic processes of
relaxation and recombination of photo-excited electrons and leads to non-vanishing shift
currents in thermal equilibrium. The goals of this methodical note is to specify and
argue the benchmarks of the BPVE theory and return the studies on the right track in the
interest of development of photovoltaic devices.

\end{abstract}

\maketitle

{\bf Introductory:} The bulk photovoltaic effect (BPVE) -- spatially uniform generation
of electric currents by light in noncentrosymmetric materials -- was investigated
massively in $70$th and $80$th of the last century. Hundreds of experimental and
theoretical papers were published and tens of different materials (ferroelectrics and
piezoelectrics) were considered. The results of these studies are summarized in
book~\cite{Book92}. They cover many aspects of the BPVE: a general definition of the
effect, its mechanisms for different types of light-induced transitions, applications to
particular materials, the influence of magnetic field, comparison between theory and
experiment, and also delusions occurred. Numerous further applications to semiconductor
nanostructures are summarized in book~\cite{IvchenkoBook}.

A new upsurge of interest in the BPVE occurred several years ago. It is caused by the
progress in material science, by increased computational possibilities, and by the
prospects of employment of the BPVE in efficient light batteries. Unfortunately, the
ongoing progress is aggravated by oversights in the basics of the BPVE theory and by
misinterpretations of the experimental and theoretical results. These drawbacks are
centered around the notion of the so-called {\em shift} current. They are relevant to
numerous recent
publications~\cite{Young12-1,Young12-2,Daranciang12,Brehm14,Wang15,Zheng15,RappeReview16,Jap1,Jap2,Liu17}.
Our goal is to outline what and why is not correct in the latest developments to return
the studies on the right track. When necessary, we refer to original papers cited
in~\cite{Book92,IvchenkoBook}.

The BPVE is conventionally defined by the tensorial relation for the DC current density
$\bm{j}$:
\begin{equation}\label{Definition}
j_i = (\beta^{L}_{inm} e_ne^*_m + \beta^C_{in} \, \kappa_n)I\,.
\end{equation}
Here $I$ is the light intensity, $\bm{e}$ is the unit polarization vector, $\bm{\kappa}
= {\rm i}(\bm{e}\times \bm{e}^*)$, and $\beta^{L}_{inm} = \beta^{L}_{imn}$ and
$\beta^C_{in}$ are two photovoltaic tensors possessing the symmetry of the piezo- and
gyration tensors, respectively. This definition employs nothing, but symmetry
considerations. The first contribution to $\bm{j}$ is nonzero for the linear
polarization ($\bm{e} = \bm{e}^*$); it corresponds to the so-called linear BPVE. The
second contribution is zero for the linear polarization and maximal for the circular
polarization ($|\bm{\kappa}| = 1$); it corresponds to the circular BPVE. Using
Eq.~(\ref{Definition}), numerous experimental data on $\bm{j}(I,\bm{e})$ were identified
with the BPVE, and nonzero components of the tensors $\hat{\beta}^{L}$ and
$\hat{\beta}^C$ were measured for tens of noncentrosymmetric materials, including
ferroelectrics BaTiO$_3$, LiNbO$_3$ and cubic piezoelectric crystals GaAs, GaP.

The microscopic theory includes general relations, simplified models, and applications
to particular materials. Most of the results are obtained within the paradigm of
quasi-free band electrons implying that the electron (hole) momentum $k$ exceeds the
reciprocal of the free-path length $\ell$. The opposite case of hopping transport is
also considered. The crystal asymmetry is presumed to be small. This means that the
asymmetry parameter $\xi_0$, defined as the ratio of non-cental ionic displacement in a
unit cell to the cell size $a$, is small, $\xi_0 \ll 1$.

{\bf Ballistic and shift currents:} A cornerstone of the microscopic BPVE theory is that
the total current density $\bm{j}$ is generally the sum of two physically different {\em
ballistic} and {\em shift} currents, $\bm{j} = \bm{j}_b + \bm{j}_{sh}$. The {\em
ballistic} current is
\begin{equation}\label{j_b}
\bm{j}_b = -e \sum\limits_{s,\bm{k}} \; f_{s,\bm{k}}\, \bm{v}_{s,\bm{k}} \;,
\end{equation}
where $e$ is the elementary charge, $\bm{v}_{s,\bm{k}} = \bm{\nabla}_{\bm{k}}\,
\varepsilon_{s,\bm{k}}/\hbar$ is the electron velocity, and $f_{s,\bm{k}}$ is the
momentum distribution in band $s$. This contribution is due to asymmetry of the momentum
distributions, $f_{s,\bm{k}} \neq f_{s,-\bm{k}}$. It can be regarded as classical one
because of the relevance to an abundance of charge-transport phenomena. The {\em shift}
current $\bm{j}_{sh}$ is caused by shifts of electrons during light-induced
transitions~\cite{Book92,IvchenkoBook,JETP82}. It originates from non-diagonal in $s$
elements of the electron-density matrix.

In noncentral media, any electronic process, including photo-excitation, recombination,
elastic and inelastic scattering, is asymmetric. Correspondingly, kinetic equations for
$f_{s,\bm{k}}$ are not inversion invariant. In the absence of thermal equilibrium, each
process contributes to~$\bm{j}_b$. However, in thermal equilibrium, where the detailed
balance between transitions ${s,\bm{k}} \hspace*{-1.2mm} \rightarrow \hspace*{-1.2mm}
{s',\bm{k}}'$ and ${s,\bm{k}} \hspace*{-1.2mm} \leftarrow \hspace*{-1.2mm} {s',\bm{k}}'$
takes place, the kinetic equations give $\bm{k}$-symmetric Fermi distributions for
electrons and holes and $\bm{j}_b = 0$. The principles of calculation of $\bm{j}_b$ are
documented for all main types of light-induced transitions including trap-band and
band-band transitions~\cite{Book92,UFN80,Phonon1,Phonon2,Coulomb}. They are illustrated
with simple models. Existence of $\bm{j}_b$ is undeniable for both the linear and
circular BPVE.

Difficulties in calculation of $\bm{j}_b$ for particular materials are rooted not only
in an insufficient knowledge of the band structure. They stem greatly from a bad
knowledge of the kinetic characteristics, such as the momentum relaxation times, the
energy relaxation times, the recombination times that involve numerous electron and
phonon bands and interaction channels. This situation is not special for the BPVE. It is
relevant to other light-induced charge transport phenomena, such as, e.g., the
photon-drag effect. Nevertheless, there are cases where $\bm{j}_b$ is calculated and
compared with experiment.

Regardless of details, the value of $\bm{j}_b$ for trap-band and band-band transitions
can be evaluated as
\begin{equation}\label{EstimateBallistic}
j_b = e\, g\, \xi_{\rm ex} \, \ell_0 \;,
\end{equation}
where $g = \alpha I/\hbar \omega$ is the generation rate, $\alpha$ is the light
absorption coefficient, $\hbar\omega$ is the light-quantum energy, $\ell_0 = v_0\tau_0$,
is the free-path length of photo-excited hot electrons (holes), $\tau_0$ is their
momentum relaxation time, and $\xi_{\rm ex}$ is the excitation asymmetry parameter. The
latter is the only BPVE-specific parameter in Eq.~(\ref{EstimateBallistic}). According
to model estimates, $\xi_{\rm ex}$ ranges from $10^{-1}$ to $10^{-3}$.

Properties of the electronic wave function $\Psi_{s,\bm{k}}(\bm{r})$ have to be
commented to avoid confusion. One might suggest that it is a Bloch function $\Psi^{\rm
B}_{s,\bm{k}}(\bm{r})$ obeying the micro-reversibility relation $\Psi^{\rm B}_{s,\bm{k}}
= \Psi^{\rm B\,*}_{s,-\bm{k}}$. With this choice, the excitation and recombination rates
for electrons are symmetric in $\bm{k}$ for the linear light polarization and
$\bm{j}_b^L = 0$. However, identification of $\Psi_{s,\bm{k}}$ with $\Psi^{\rm
B}_{s,\bm{k}}$ is generally incorrect. For trap-band transitions, the effect of trap
potential is important. Here, the functions $\Psi_{\bm{k}}^-(\bm{r})$ and
$\Psi_{\bm{k}}^+(\bm{r})$ comprising converging ($-$) and diverging ($+$) waves, see
Fig.~1a, must be used to describe the excitation and recombination
probabilities~\cite{LandauQM}.
\begin{figure}[h]
\centering
\includegraphics[width=8.3cm]{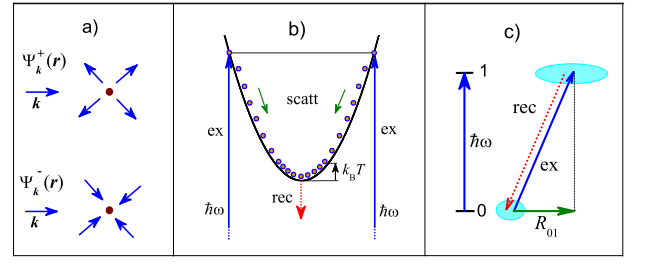}
\caption{a) The functions $\Psi^+_{\bm{k}}(\bm{r})$ and $\Psi^-_{\bm{k}}(\bm{r})$
comprising the diverging and converging waves and relevant to the recombination and
excitation processes, respectively. b) A cycle of excitation, energy relaxation,
thermalization, and recombination processes typical of trap-band and band-band
transitions. c) Zero steady-state shift current for transitions between two localized
states $0$ and $1$. }\label{Processes}
\end{figure}
None of the $\Psi_{\bm{k}}^{\pm}$ functions obeys the micro-reversibility relation; this
gives asymmetry of the excitation and recombination rates. At the same time, the
micro-reversibility relation $\Psi_{-\bm{k}}^{+\,*} = \Psi_{\bm{k}}^{-}$ links the
differential probabilities of excitation and recombination: $w^{\rm ex}_{\bm{k}}(\bm{e})
= w^{\rm rec}_{-\bm{k}}(\bm{e}^*)$~\cite{Book92}. An analogous situation takes place for
VB$\to$CB transitions. The Coulomb interaction of light-induced electrons and holes,
which is typically strong, causes not only localized excitonic states, but also
inequality $\Psi_{s,\bm{k}} \neq \Psi_{s,-\bm{k}}^*$~\cite{Book92,Coulomb}. The
excitation asymmetry parameter can be estimated as $\xi_{\rm ex} \approx \xi_0$. For
transitions between electronic (or hole) bands, the electron-phonon interaction plays
the role of perturbing potential~\cite{Book92,Phonon1,Phonon2}. Here the excitation
asymmetry parameter can be estimated as $\xi_{\rm ex} \approx \xi_0 a/\ell_0 \ll \xi_0$.

The terms {\em shift} BPVE and $\bm{j}_{sh}$ were introduced in~1982 at the breakthrough
in understanding of the physics of the non-diagonal contribution to
$\bm{j}$~\cite{JETP82}. It was realized that an electronic transition from Bloch state
$|n\rangle = |s,\bm{k}\rangle$ to Bloch state $|n'\rangle = |s',\bm{k}'\rangle$ is
accompanied by the shift $\bm{R}_{n'n}$ within a unit crystal cell:
\begin{equation}\label{Shift}
\bm{R}_{n'n} = -\bm{R}_{nn'} = - (\bm{\nabla}_{\bm{k}} + \bm{\nabla}_{\bm{k}'})\,
\Phi_{n'n} + \bm{\Omega}_n - \bm{\Omega}_{n'} \,,
\end{equation}
where $\Phi_{n'n}$ is the phase of the transition matrix element and $\bm{\Omega}_n$ is
the Berry connection as defined in~\cite{LandauSP}. The shift is even in $\bm{k}$ and
$\bm{k}'$ and nonzero only in the absence of inversion symmetry. It does not depend on
the choice of the phase of $\Psi^B_{s,\bm{k}}$ (the gauge invariance). The shift current
is expressed generally by $\bm{R}_{n'n}$ as~\cite{Book92,IvchenkoBook,JETP82}
\begin{equation}\label{j_sh}
\bm{j}_{sh} = e \sum\limits_{n,n'} \, W_{n'n}\,\bm{R}_{n'n},
\end{equation}
where $W_{n'n}(f_n,f_{n'})$ is the transition rate from $n$ to $n'$. In thermal
equilibrium, where the detailed balance takes place, $\bm{j}_{sh} = 0$. Since
$\bm{R}_{n'n}$ already accounts for the crystal asymmetry, $\bm{j}_{sh}$ has to be
calculated on symmetric distributions, $f_{s,\bm{k}} = f_{s,-\bm{k}}$. Small differences
between $\Psi^B_{s,\bm{k}}$ and $\Psi_{s,\bm{k}}$ are not important for the shift BPVE.
%The $\bm{\Omega}$ contribution to $\bm{j}_{sh}$ is zero in steady state.

Application of Eqs.~(\ref{Shift},\,\ref{j_sh}) to light-induced band-band transitions
gives an explicit expression for $\bm{j}_{sh}$~\cite{Book92,JETP82}. It consists of
partially compensating contributions $\bm{j}^{\rm ex}_{sh}$ and $\bm{j}^{\rm rec}_{sh}$
relevant to electron shifts during the excitation and recombination processes. The
excitation contribution $\bm{j}^{\rm ex}_{sh}$ is proportional to the light intensity
$I$ and includes no kinetic characteristics. The recombination contribution $\bm{j}^{\rm
rec}_{sh}$ incorporates such characteristics: Hot electrons (holes) experience the
energy relaxation and recombine being predominantly (or partially) thermalized, see
Fig.~1b. Since the rates of excitation and recombination are the same in steady state,
the value of $\bm{j}_{sh}$ depends on particular values of the shift for hot and
thermalized electrons. In the absence of the relaxation, as in the case of Fig.~1c, the
total shift current would be zero because of an exact compensation of $\bm{j}_{sh}^{\rm
ex}$ and $\bm{j}_{sh}^{\rm rec}$. Simple models show that the contribution $\bm{j}^{\rm
rec}_{sh}$ can be dominating in ferroelectrics~\cite{Book92,Ferro88}. The value of the
shift current can be estimated as
\begin{equation}\label{EstimateShift}
j_{sh} = e\, g\, \bar{R} \,,
\end{equation}
where $\bar{R} \approx \xi_0a$ is an effective shift.

The condition of dominating ballistic current is thus $\xi_{ex}\ell_0 \gg \bar{R}$. It
is expected to be fulfilled for VB$\to$CB transitions where the strong electron-hole
interaction provides the excitation asymmetry parameter $\xi_{ex} \approx \xi_0$. In the
case of transition between electronic (or hole) bands we have $\xi_{ex} \approx
a/\ell_0$ and $j_b \approx j_{sh}$. More accurate calculations support these simple
estimates, see also below.

{\bf An alternative BPVE theory:} Works~\cite{Young12-1,Young12-2,RappeReview16} put
forward an alternative version of the linear BPVE theory which strongly contradicts
to~\cite{Book92,IvchenkoBook}. This version is centered around the notion of shift
current and deals with band-band transitions. It treats the linear BPVE as a dynamic
nonlinear-optical phenomenon, free of influence of kinetic parameters. The classical
ballistic contribution $\bm{j}_b$, caused by asymmetry of the momentum distributions for
electrons and holes, is absent within this approach. The basic relation for
$\bm{j}_{sh}$ is not different here from the expression of~\cite{Book92,JETP82} for
$\bm{j}^{\rm ex}_{sh}$. In other words, the shift currents considered do not include the
recombination contribution~$\bm{j}^{\rm rec}_{sh}$. Since the expression for
$\bm{j}^{\rm ex}_{sh}$ includes only band (but not kinetic) characteristics, it can be
used for calculations from "first principles". Such calculations have been performed for
a number of ferroelectric
materials~\cite{Young12-1,Young12-2,Daranciang12,Brehm14,Wang15,Zheng15,Liu17}. Broad
frequency ranges, which are not restricted to the excitation of charge carriers near
band edges, is a remarkable feature of these calculations. Neither discussion nor
criticism of~\cite{Book92} can be found within the alternative BPVE theory.

Remarkably, the formulae for $\bm{j}$ of~\cite{Baltz81,Sipe00} served as the basis for
the new developments. The model employed in these papers includes only the interaction
of Bloch electrons with a classical electromagnetic field. The electron-phonon and
electron-hole interactions leading to the ballistic current $\bm{j}_b$, as well as the
common processes of energy relaxation, thermalization, and recombination, are beyond
this oversimplified model. The authors of~\cite{Sipe00} and of the computational
papers~\cite{Young12-1,Young12-2,Daranciang12,Brehm14,Wang15,Zheng15,Liu17} refer
to~\cite{Book92} as to a general source, but pay no attention either to the specific
results on $\bm{j}_b$ and $\bm{j}_{sh}$ or the general argumentation. In essence, the
model expression for $\bm{j}^{L}$ of~\cite{Baltz81,Sipe00} is considered as omnipotent
one for the linear BPVE. Referring to~\cite{Young12-1,Young12-2,Baltz81,Sipe00},
numerous experimental
papers~\cite{Cote02,Bieler07,Loata08,Racu09,Kohli11,Somma14,Ogava17} claim for
observation and employment of the shift currents.

Surprizingly, no attention was paid to numerous evidences (qualitative, quantitative,
experimental) of the presence of the ballistic contributions to $\bm{j}^L$ as well as to
pathology of the model ignoring the energy relaxation and recombination processes.
Ignorance of the recombination leads evidently to a wrong conclusion about the presence
of steady-state shift current for an ensemble of localized two-level centers subjected
to the resonant excitation, see Fig.~1c. For crystals of the pyroelectric symmetry, it
leads to nonzero currents in thermal equilibrium owing to the thermal-photon excitation
processes, i.e., to perpetual motion machine of the second kind.

{\bf Applications to GaAs crystals:} The BPVE theory was applied to a number of
noncentrosymmetric semiconductors with well known band structure and explored kinetic
properties (GaAs, GaP, Te). The results of the calculations, performed for
near-band-edge absorption, have shown a good agreement with experimental
data~\cite{Book92}. Altogether, they evidence that the ballistic current $\bm{j}_b$ is
either dominating over or comparable with~$\bm{j}_{sh}$. This refers both to the linear
and circular BPVE.

Here we exhibit some results obtained for GaAs. This cubic piezoelectric possesses only
the linear BPVE characterized by the component $\beta = \beta^L_{123}$ of the
photovoltaic tensor $\hat{\beta}^L$. The band structure of GaAs for $ka \ll 1$ is shown
in Fig.~1a. It is characterized by a single CB and a compound VB comprising
\begin{figure}[h]
\centering \vspace*{2mm}
\includegraphics[width=8cm]{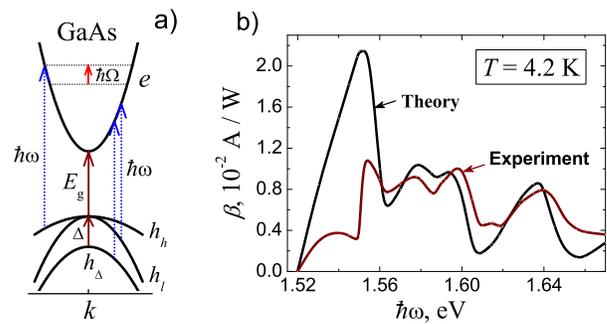}
\caption{a) Band structure of GaAs and the main light-excitation channels (blue arrows).
b) Spectral dependence of the BPVE-tensor element $\beta$, theory and
experiment~\cite{Book92,GaAsExperiment}.}\label{GaAs}
\end{figure}
the light ($h_l$), heavy ($h_h$), and split-off ($h_{\Delta}$) hole sub-bands. The
bandgap $E_g$ and spin-orbit splitting $\Delta$ are $\simeq 1.53$ and $0.34$~eV,
respectively. The band structure is described by the Kane Hamiltonian which is a
$\bm{k}$-dependent $8 \times 8$ matrix~\cite{PikusBook}. Calculations of
$\bm{j}_b(\omega)$ for CB$\to$VB transitions were performed
in~\cite{GaAsTheory,PGEsGaAs82,GaAsExperiment} near the fundamental absorption edge,
$E_g < \hbar \omega < E_g + \Delta$ for He-temperatures. The Coulomb interaction between
photo-excited electrons and holes was found to be the dominating mechanism of the
excitation asymmetry. Complexity of the quantitative calculations is rooted not only in
simultaneous excitation of three bands, but also in involvement of different mechanisms
of momentum relaxation. The strongest of them is scattering on optical phonons. This
mechanism is allowed for sufficiently large distances $\hbar \omega - E_g$ and gives
only partial momentum relaxation. This is why weaker kinetic processes -- scattering on
acoustic phonons and remnant neutral and charged  point defects -- contribute to the
relaxation. As the result, the dependence $\beta(\hbar\omega)$ acquires a characteristic
jagged structure, see Fig.~\ref{GaAs}b, which is the fingerprint of the ballistic nature
of $\bm{j}$. The shift current $j_{sh}$ was found to be smaller than $j_b$ by more than
one order of magnitude.

BPVE experiments were performed with epitaxial films of GaAs at $T =
4.2$~K~\cite{GaAsExperiment}; they included also Hall measurements. The main features of
the experimental spectrum in Fig.~2b nicely correspond to the theory (to the spectral
dependence of the free-pass length $l_0$). The maximal values of the Hall mobility and
free-pass length occur at $\simeq 1.56$~eV; they are about $5 \times 10^5$~cm$^2$/Vs and
$8\,\mu$m. The corresponding excitation-asymmetry parameter is $\xi^{\max}_{\rm ex}
\approx 2 \times 10^{-3}$.

Importantly, theoretical and experimental studies of two allied kinetic photovoltaic
phenomena -- the photon-drag effect and the surface photovoltaic effect -- were
performed independently for the same interband transitions in GaAs at Helium
temperatures~\cite{Drag81,Drag82,SurfacePGE}. A good agreement between complicated
theoretical and experimental spectral dependences was obtained. A great body of results
gives thus a comprehensive knowledge of the interband photovoltaic effects in GaAs.

Relationship between $j_b$ and $j_{sh}$ changes dramatically for transitions between the
$h_l$ and $h_h$ hole sub-bands when the Coulomb mechanism of the ballistic current is
absent. In particular, it was shown theoretically and experimentally for $\hbar\omega =
117$~meV that a strong compensation of $\bm{j}_b$ and $\bm{j}_{sh}$ takes place in the
temperature range $(130 - 570)\,$K~\cite{Book92,GaAsFTT84,GaAsFTT85}. The total current
$\bm{j} = \bm{j}_b + \bm{j}_{sh}$ is relatively small and sign-changing.

{\bf Influence of magnetic field:} Investigation of the influence of magnetic field on
the photovoltaic current sheds light on the nature of the shift currents and on close
links between $\bm{j}_b$ and $\bm{j}_{sh}$. Since the electronic shifts occur
instantaneously during the light-induced transitions, one might suggest that the effect
of magnetic field $\bm{H}$ on $\bm{j}_{sh}$ is negligible compared to the effect on
$\bm{j}_b$. The latter can be viewed for small fields as generation of the Hall current
$\delta \bm{j}_b \approx \pm (\mu_0/c)(\bm{H} \times \bm{j}_b)$, where $\mu_0$ is the
mobility of the dominating photo-excited charge carriers, $c$ is the speed of light, and
the signs $+$ and $-$ correspond to dominating electrons and holes. This effect is due
to the action of the Lorentz force during the free electron path. Applying large
magnetic fields, $\mu_0H/c \gg 1$, leading to suppression of the transverse (to
$\bm{H}$) ballistic current, one can hope to get experimentally the transverse shift
current.

Unfortunately, the above suggestion about a negligible influence of the magnetic field
on $\bm{j}_{sh}$ is not fully correct. The point is that this field breaks the
time-reversal invariance, so that $\Psi^B_{\bm{k}} \neq \Psi^{B*}_{-\bm{k}}$ for Bloch
electrons. This causes magneto-induced asymmetry of photo-excitation (the inequality
$w^{\rm ex}_{\bm{k}} \neq w^{\rm ex}_{-\bm{k}}$) without any perturbing effect of the
electron-hole and electron-phonon interactions. Correspondingly, an additional
magneto-induced contribution $\delta \bm{j}^{\rm M}_b$ to the ballistic current joins
the game. Surprizingly, this ballistic contribution is expressed by the shift
current~\cite{Book92,MagnetoInduced84}: $\delta \bm{j}^{\rm M}_b = \pm (C\mu_0/c)(\bm{H}
\times \bm{j}_{sh})$, where $C$ is a model-dependent dimensionless constant whose value
can be smaller or modestly larger than $1$. This formula shows that the shift and
ballistic currents become mutually related in the presence of magnetic field. For the
large magnetic fields, the current $\delta \bm{j}^{\rm M}_b$ compensates the transverse
component of the shift current, so that the total transverse current vanishes. Thus, a
decisive phenomenological separation of $\bm{j}_b$ and $\bm{j}_{sh}$ is not possible in
Hall experiments. With simplifying assumptions, one can demonstrate merely that
$\bm{j}_{sh}$ is present.

Employment of the Hall model for determination of the mobility and sign of the
photo-excited electrons is justified only in the cases when $j_b \gg j_{sh}$, including
the VB$\to$CB and trap-band transitions. Large values of the mobility obtained in such
experiments~\cite{Book92,GaAsExperiment,Hall1,Hall2,Hall3}, $\mu_0 \gtrsim
10^2$~cm$^2$/Vs, support once again the conclusion about smallness of the shift current
and evidence in favor of the ballistic models of the BPVE.

{\bf Summary:} There are no reasons to identify the measurable total BPVE current
$\bm{j}$ or the linear BPVE current $\bm{j}^L$ with the shift current $\bm{j}_{sh}$.
Numerous theoretical and experimental arguments indicate that the classical ballistic
contribution $\bm{j}_b$, caused by asymmetry of the momentum distributions for electrons
and/or holes, is either dominating or substantial. The ballistic current is missed in
within the alternative BPVE theory. The recombination contribution $\bm{j}^{\rm
rec}_{sh}$ to the shift current, which is nonzero in ferroelectric crystals, is also
missed within the alternative theory. The latter leads to perpetual motion machine of
the second kind.

Both missed contributions, $\bm{j}_b$ and $\bm{j}^{\rm rec}_{sh}$, involve kinetic
characteristics, such as momentum and energy relaxation times. Correspondingly, it is
necessary to treat the BPVE as a kinetic phenomenon, substantially different from and
more complicated than the quadratic-response effects caused solely by the band
structure. The difference stems from the fact that generation of a DC current is always
a dissipative process in contrast to generation of the nonlinear polarization. This
difference disappears in the frequency domain when the frequency difference between two
light waves substantially exceeds the reciprocal of the momentum relaxation time
$\tau_0$.

The shift and ballistic contributions to the BPVE current possess the same symmetry
properties, they cannot be decisively separated in light-polarization experiments. The
same is valid greatly with respect to Hall measurements. Deep knowledge of the
electronic band structure and kinetic processes is necessary to judge about the
relationship between $\bm{j}_b$ and $\bm{j}_{sh}$.

\vspace*{2mm} Accepted to Physics-Uspekhi, Advances in Physical Sciences, DOI:
10.3367/UFNe.2019.06.038578.

\end{document}